\def\gp{\gamma_+}
\def\gm{\gamma_-}
\def\bm{\beta_-}
\def\bp{\beta_+}
\def\gt{\gamma_t}
\def\gcm{\gamma_{cm}}

\def\bcm{\beta_{cm}}

\def\ugr{\lower4pt \hbox{$\buildrel > \over \sim$}}
\def\ukl{\lower4pt \hbox{$\buildrel < \over \sim$}}

\def\bg{\beta_{\Gamma}}

\def\gpm{\gamma_{\pm}}

\def\bpm{\beta_{\pm}}

\def\gepm{\gamma_{1\pm}}

\def\gzpm{\gamma_{2\pm}}

\def\ecm{\epsilon_{cm}}

\def\emp{$e^-$-$e^+$}

\def\st{\sigma_T}

\def\gamr{gamma-ray }
\def\eph{\epsilon_{ph}}

\documentstyle[epsf, rotate]{laa}
\begin{document}

\thesaurus{02.16.1, 02.18.1, 02.18.2, 11.10.1 13.07.3}
\title{$\gamma$-ray emission and spectral evolution of pair plasmas in AGN jets}
\subtitle{I. General theory and a prediction for the GeV  -- TeV emission 
from ultrarelativistic jets}
\author{M.  B\"ottcher \and H. Mause \and R. Schlickeiser}
\institute{Max-Planck-Institut f\"ur Radioastronomie, Postfach 20 24, 53
010 Bonn, Germany}
\date{Received 25 March 1996; Accepted 30 January 1997}
\offprints{M. B\"ottcher}

\maketitle

\begin{abstract}
The evolution of the particle distribution functions inside a
relativistic jet containing an \emp pair plasma and of the
resulting $\gamma$-ray and X-ray spectra is investigated.
The first phase of this evolution is governed by heavy radiative
energy losses. For this phase, approximative expressions for
the energy-loss rates due to inverse-Compton scattering, using the
full Klein-Nishina cross section, are given as one-dimensional
integrals for both cooling by inverse-Compton scattering of synchrotron
photons (SSC) and of accretion disk photons (EIC).

We calculate instantaneous and time-integrated $\gamma$-ray spectra 
emitted by such a jet for various sets of parameters, deducing
some general implications on the observable broadband radiation.
Finally, we present model fits to the broadband spectrum of 
Mrk~421. We find that the most plausible way to explain both 
the quiescent and the flaring state of Mrk~421 consists of
a model where EIC and SSC dominate the observed spectrum in 
different frequency bands. In our model the flaring state 
is primarily related to an increase of the maximum Lorentz 
factor of the injected pairs.

\end{abstract}

\section{Introduction}
Accretion of matter onto a central black hole is the most relevant 
process to power active galactic nuclei (Lynden-Bell 1969, Salpeter 
1969, Rees 1984). However, the details of the conversion processes of 
gravitational energy into observable electromagnetic radiation are 
still largely unknown. The discovery of many blazar-type AGNs 
(Hartman et al. 1992, Fichtel et al. 1993) as sources of high-energy 
\gamr radiation dominating the apparent luminosity, has revealed 
that the formation of relativistic jets and the acceleration of 
energetic charged particles, which generate nonthermal radiation,
are key processes to understand the energy conversion process. 
Emission from relativistically moving sources is required to 
overcome \gamr transparency problems implied by the measured large 
luminosities and short time variabilities (for review see Dermer 
\& Gehrels 1995).

Repeated \gamr observations of AGN sources have indicated a typical 
duty cycle of \gamr hard blazars of about 5 percent, supporting 
a ``2-phase" model for the central regions of AGNs (Achatz 
et al. 1990, Schlickeiser \& Achatz 1992). According to the 
2-phase model the central powerhouse of AGNs undergoes two repeating 
phases: in a ``quiescent phase" over most of the time ($\sim $95 
percent) relativistic charged particles are efficiently accelerated 
in the central plasma near the black hole, whereas in a short
and violent ``flaring phase" the accelerated particles are ejected 
in the form of plasma blobs along an existing jet structure.

We consider the acceleration of charged particles during the 
quiescent phase. The central object accretes the surrounding 
matter. Associated with the accretion flow is low-frequency 
magnetohydrodynamic turbulence which is generated by various 
processes as e.g.:

\noindent (a) turbulence generated by the rotating accretion disk
at large eddies and cascading to smaller scales (Galeev et al. 1979);

\noindent (b) stellar winds from solar-type stars in the central star 
cluster deliver plasma waves to the accretion flow;

\noindent (c) infalling neutral accretion matter becomes ionized
by the ultraviolet and soft X-ray radiation of the disk. These 
pick-up ions in the accretion flow generate plasma waves by virtue 
of their streaming (Lee \& Ip 1987);

\noindent (d) if standing shocks form in the neighbourhood of the 
central object they amplify any incoming upstream turbulence in the 
downstream accretion shock magnetosheath (McKenzie \& Westphal 1969, 
Campeanu \& Schlickeiser 1992).

These low-frequency MHD plasma waves from the accretion flow are the 
source of free energy and lead to stochastic acceleration of charged 
particles out of the thermal accretion plasma.

The dynamics of energetic charged particles (cosmic rays) in cosmic 
plasmas is determined by their mutual interaction and interactions 
with ambient electromagnetic, photon and matter fields. Among these 
by far quickest is the particle-wave interaction with electromagnetic 
fields, which very often can be separated into a leading field 
structure $F_o$ and superposed fluctuating fields $\delta F$.
Theoretical descriptions of the transport and acceleration of cosmic 
rays in cosmic plasmas are usually based on transport equations which 
are derived from the Boltzmann-Vlasov equation into which the 
electromagnetic fields of the medium enter by the Lorentz force term.
The quasilinear approach to wave-particle interaction is a second-order 
perturbation approach in the ratio $q_L\equiv ({\delta F/F_o})^2$ 
and requires smallness of this ratio with respect to unity. In most 
cosmic plasmas this is well satisfied as has been established either
by direct in-situ electromagnetic turbulence measurements in 
interplanetary plasmas, or by saturation effects in the growth of
fluctuating fields. Nonlinear wave-wave interaction rates and/or 
nonlinear Landau damping set in only at appreciable levels of 
$(\delta F)^2$ and thus limit the value of $q_L\le 1$. We assume 
the AGN plasma to have very high conductivity so that any 
large-scale steady electric fields are absent. We then consider 
the behaviour of energetic charged particles in a uniform magnetic 
field with superposed small-amplitude $(\delta B)^2 \ll B_o^2$
plasma turbulence ($\delta \vec{E}, \delta \vec{B}$) by calculating 
the quasilinear cosmic ray particle acceleration rates and transport 
parameters. This is by no means trivial since especially for the 
interaction of non-relativistic charged particles with ion- and 
electron-cyclotron waves thermal resonance broadening effects are 
particularly important (Schlickeiser \& Achatz 1993, Schlickeiser 
1994). The acceleration rates and spatial transport parameters are 
then used in the kinetic diffusion-convection equation for the 
isotropic part of the phase space density of charged particles
$F(x,p,t)$ which for non-relativistic bulk speed $u \ll c$ reads

$${{\partial F}\over {\partial t}}-\, S_o= {\partial \over {\partial 
x}}[\kappa {{\partial F}\over {\partial x}}] 
$$
\begin{equation}
\;\;\; - \, u{{\partial F}\over {\partial x}}+{p\over 3}{{\partial 
u}\over {\partial x}} {{\partial F}\over {\partial p}}+\,{1\over 
p^2}{\partial \over {\partial p}} [p^2A{ {\partial F}\over {\partial 
p}} -p^2 \dot{p}_{\rm loss} F].
\end{equation}
Here $x$ denotes the spatial coordinate along the ordered magnetic 
field, $p$ the cosmic ray particle momentum, $\kappa $ is the spatial 
diffusion coefficient, $A$ the momentum diffusion coefficient, and 
$S_o$ denotes the "Stossterm" describing the mutual interaction of 
the charged particles and their injection.

With respect to the generation of energetic charged particles, the 
basic transport equation (1) shows that stochastic acceleration of 
particles, characterized by the acceleration time scale $t_A=p^2/A$, 
competes with continuous energy loss processes $\dot{p}_{\rm loss}$, 
characterized by energy loss time scales $t_L=p/|\dot{p}_{\rm loss}|$.
Dermer et al. (1996) have recently inspected the acceleration 
of energetic electrons and protons in the central AGN plasma by 
comparing the time scales for stochastic acceleration with the
relevant energy loss time scales. At small proton momenta the 
Coulomb loss time scale is extremely sensitive to the background 
plasma density and temperature, and for slight changes in the 
values of these parameters cosmic ray protons may not be 
accelerated above the Coulomb barrier. Although at small particle 
momenta the plasma wave's dissipation and the interaction with the 
cyclotron waves become decisive and might modify the acceleration 
time significantly, the results of Dermer et al. (1996)
demonstrate that reasonable central AGN plasma parameter values 
are possible where the low-frequency turbulence energizes protons 
to TeV and PeV energies where photo-pair and photo-pion production 
are effective in halting the acceleration (Sikora et al. 1987,
Mannheim \& Biermann 1992). According to the results 
of Dermer et al. (1996) it takes about $\simeq 10M_8$ days for 
the protons to reach these energies, where $M_8$ is the mass of the 
central black hole in units of $10^8 \, M_{\odot}$. The corresponding
analysis for cosmic ray electrons shows that the external compactness 
provided by the accretion disk photons (Becker \& Kafatos 1995) 
leads to heavy inverse Compton losses which suppress the acceleration 
of low-energy electrons beyond Lorentz factors of $\gamma 
\approx 10-100$. It seems that due to their much smaller radiation 
loss rate cosmic ray protons are effectively accelerated during the 
quiescent phase in contrast to low energy electrons.

Now an important point has to be emphasized: {\it once the accelerated 
protons reach the thresholds for photo-pair ($\gamma_{p,th}
= m_ec^2/<\epsilon >=5\cdot 10^4\epsilon^{-1}_1$) and photo-pion
production and the threshold for pion production in inelastic 
proton-matter collisions they will generate plenty of secondary 
electrons and positrons of ultrahigh energy} which are now 
injected at high energies ($E_s\ge 25 {\rm GeV} \epsilon ^{-1}_1$)
into this acceleration scheme. $<\epsilon > = 10 \epsilon_1 $ eV 
denotes the mean accretion disk photon energy. It is now of 
considerable interest to follow the evolution of these injected
secondary particles.

Although many details of this evolution are poorly understood, it is 
evident that the further fate of the secondary particles depends 
strongly on whether they find themselves in a compact environment 
set up by the external accretion disk, or not. As has been pointed 
out by Dermer \& Schlickeiser (1993b) as well as Becker \& Kafatos 
(1995) the size of the \gamr photosphere (where the compactness is 
greater unity so that any produced \gamr photon is pair-absorbed) is 
strongly photon energy dependent. The \gamr photosphere attains 
its largest size at photon energies $E_p\simeq 50 \epsilon_1^{-1}$
GeV. Secondary particles within the photosphere having energies 
$E_s\le E_p$ will initiate a rapid electromagnetic cascade which 
has been studied by e.g. Mastichiadis \& Kirk (1995), which might
even lead to runaway pair production and associated strong X-ray 
flares (Kirk \& Mastichiadis 1992), and/or due to the violent 
effect of a pair catastrophy (Henri \& Pelletier 1993) ultimately 
lead to an explosive event and the emergence of a relativistically 
moving component filled with energetic electron-positron pairs.

In contrast, if the secondary particles are generated outside the 
\gamr photosphere the secondary electrons and positrons will quickly 
cool by the strong inverse Compton losses generating plenty of
\gamr emission up to TeV energies. As solutions of the electron 
and positron transport equation (1) for this case demonstrate 
(Schlickeiser 1984, Pohl et al. 1992) a cooling particle 
distribution ($N(p) \propto p^{-2}$) with a strong cutoff 
at low (but still relativistic) momentum $p_c$ develops, 
which grows with time as more and more protons hit the 
photo-pair and photo-pion thresholds. Such bump-on-tail 
particle distribution functions, which are inverted 
($\partial F/\partial p>0$) below $p_c$, are collectively unstable 
with respect to the excitation of electromagnetic and electrostatic 
waves such as oblique longitudinal Langmuir waves (Lesch 
et al. 1989). As described in detail by Lesch \& Schlickeiser 
(1987) and Achatz et al. (1990), depending on 
the local plasma parameters (mainly the electron temperature of the 
background gas and the density ratio $N_o/n_e$ of relativistic 
electrons and positrons to thermal electrons) these Langmuir waves 
either (1) lead to quasilinear plateauing of the inverted distribution 
function and rapid collective thermalization of the electrons and 
positrons, or (2) are first damped by nonlinear Landau damping, but 
ultimately heat the background plasma strongly via the modulation 
instability once a critical energy density in Langmuir waves $W_c = 
n_e(k_BT_e/m_ec^2)^2$ has been built up. Both relaxation mechanisms 
terminate the quiescent phase of the acceleration process. The 
almost instantaneous increase of the background gas entropy due to
the rapid modulation instability heating again leads to an explosive
outward motion of the plasma blob carrying the relativistic particles 
away from the central object.

As we have discussed, in both cases it is very likely that at the end 
of the quiescent phase an explosive event occurs that gives rise to 
the emergence of a new relativistically moving component filled with 
energetic electron-positron pairs. It marks the start of the flaring 
phase in \gamr blazars. The initial starting height of the emerging 
blob $z_i$ entering the calculation of the \gamr flux should be closely 
related to the size of the acceleration volume in the quiescent phase
mainly determined by the maximum size of the \gamr photosphere.
This scenario is supported by the measurements of Babadzanhanyants 
\& Belokon (1985) that in 3C 345 and other quasars optical bursts 
are in close time correlation with the generation of compact radio 
jets. A similar behaviour has also been observed during the recent 
simultaneous multiwavelength campaign on 3C~279 (Hartman et al. 1996).
Further corroborative evidence for this scenario is provided 
by the recent discovery of superluminal motion components in the 
\gamr blazars PKS~0528+134 (Pohl et al. 1995) and PKS~1633+382 
(Barthel et al. 1995) that demonstrate a close physical connection 
between \gamr flaring and the ejection of new superluminal jet 
components in blazars.

It is the purpose of the present investigation to follow the time 
evolution of the relativistic electrons and positrons as the 
emerging relativistic blob moves out. Because of the very short 
radiative energy loss time scales of the radiating electrons and 
positrons it is important to treat the spectral evolution of the 
radiating particles self-consistently. In earlier work Dermer \& 
Schlickeiser (1993a) and Dermer et al. (1997) 
have studied the spectral time evolution from a modified Kardashev 
(1962) approach by injecting instantaneously a power law electron and 
positron energy spectrum at height $z_i$ at the beginning of the 
flare, and calculating its modification with height in the 
relativistically outflowing blob due to the operation of various 
continous energy loss processes as inverse Compton scattering, 
synchrotron radiation, nonthermal bremsstrahlung emission and Coulomb 
energy losses. Here we generalize their approach by accounting
for Klein-Nishina effects and including as well external 
inverse-Compton scattering as synchrotron self-Compton scattering 
self-consistently.

The acceleration scenario described above leads us to the
following assumptions on the distribution of pairs inside
a new jet component at the time of their injection: If
the pairs are created by photo-pair production, their minimum
Lorentz factor is expected to be in the range of the threshold
value of the protons' Lorentz factor for photo-pair production.
We use the standard accretion disk model by Shakura \& Sunyaev
(1973), which we will describe in more detail in the next
section, to fix this threshold, determined by the average 
disk photon energy $\langle \epsilon \rangle$. The 
pair distribution above this cutoff basically reflects the 
acceleration spectrum of the protons, i. e. a power-law distribution 
with spectral index $2 \ukl s \ukl 3$ ($n(\gamma) \sim \gamma^{-s}$)
which extends up to $\gamma_{2\pm} \sim 10^6$. This yields the initial 
pair distribution functions

\begin{equation}
$$ f_{\pm} (\gpm) = \gpm^{-\sigma} \hskip 2cm \gepm \le \gpm
\le \gzpm $$
\end{equation}
with $4 \ukl \sigma \ukl 5$. The differential number of particles 
in the energy intervall $[\gpm, \gpm + d\gpm]$ per unit volume
is then given by

\begin{equation}
d \, n_{\pm} = 4 \, \pi \, N_{\pm} \, \gpm^2 \bpm \, 
f_{\pm} (\gpm) d\gpm
\end{equation}
where $N_{\pm}$ is a normalization factor related to the total particle 
density $n_{\pm}$ through $N_{\pm} \approx n_{\pm} {\sigma - 3 \over 
4 \, \pi} \gepm^{\sigma - 3}$. 

The detection of TeV $\gamma$-rays from Mrk~421 suggests that
such components must be produced/accelerated outisde the
$\gamma$-ray photosphere for photons of energy $\sim 1$ TeV.
The height of this photosphere due to the interaction of
$\gamma$-rays with accretion disk radiation will be determined 
self-consistently in section 4. Backscattering of accretion disk 
radiation by surrounding clouds is negligible in the case of BL Lac 
objects emitting TeV $\gamma$-rays (B\"ottcher \& Dermer, 1995). 

We point out that most of our basic conclusions are also 
valid if the pairs inside the blob are accelerated by
other mechanisms, e. g. by a relativistic shock propagating
through the jet.

Beginning at the injection height (henceforth denoted as $z_i$), 
we follow the further evolution of the pair distribution 
and calculate the emerging photon spectra.

\begin{figure}
\epsfxsize=6cm
\epsffile[20 150 450 650] {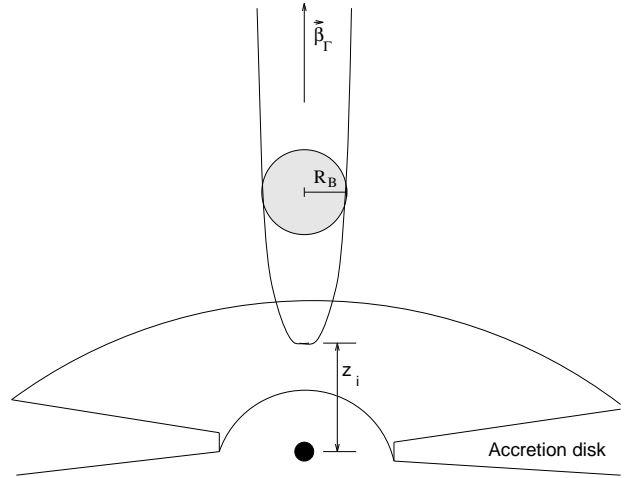}
\caption[]{Model for the geometry of a relativistic AGN jet}
\end{figure}

The negligibility of pair absorption due to the interaction 
with the synchrotron and $\gamma$-ray emission from the jet 
is checked self-consistently during our calculations.

Interactions of the jet pair plasma with dilute surrounding material
will cause turbulent Alfv\'en and Whistler waves. It has been
shown by Achatz \& Schlickeiser (1993) that a low-density, 
relativistic pair jet is rapidly disrupted as a consequence of
the excitation of such waves. Thus, to insure stability of the
beam over a sufficient length scale, we need that the density of
pairs inside the jet exceeds the density of the surrounding
material. In this case, pitch angle scattering on plasma 
wave turbulences leads to an efficient isotropization of
the momenta of the pairs in the jet without destroying the
jet structure. Thus, additional assumptions on our initial
conditions are that the particle momenta are isotropically
distributed in the rest frame of a new jet component (blob)
and that the density of pairs in the jet $n_j \gg n_b$ where
$n_b$ is the density of the background material.

This study is devided into two papers. In the first (present)
paper we investigate the details of electron/positron cooling due 
to inverse-Compton scattering and follow the pair distribution and 
photon spectra evolution during the first phase in which the 
system is dominated by heavy radiative losses. In the second
paper we will consider the later phase of the evolution
where collisional effects (possibly leading to thermalization)
and reacceleration become important, and a plausible model for
MeV blazars which follows directly from our treatment will be
presented (B\"ottcher, Pohl \& Schlickeiser, in prep.).

In section 2 of this first paper, we describe in detail 
how to calculate the energy-loss rates due to the 
various processes which we take into account and
give useful approximative expressions for the inverse-Compton
losses (as well scattering of accretion disk radiation as of
synchrotron radiation), including all Klein-Nishina effects.
In section 3, we discuss the relative importance of the various
processes. The location of the $\gamma$-ray photosphere
for TeV $\gamma$-rays is briefly outlined in section 4.
The technique used to follow the evolution of the pair 
distributions is described in section 5. In section 6, we describe 
how to use the pair distributions resulting from our simulations in 
order to calculate the emanating $\gamma$-ray spectra, and in 
section 7 we discuss general results of our simulations giving 
a prediction for GeV -- TeV emission from $\gamma$-ray blazars 
which due to the lack of sensitivity of present-day instruments 
in this energy range could not be observed until now. Only two 
extragalactic objects have been detected as sources 
of TeV emission, namely Mrk~421 (Punch et al. 1992) 
and Mrk~501 (Quinn et al. 1995). In Section 8, we use our
code to fit the observational results on the broadband
emission during the TeV flare of Mrk~421 in May 1994 and
on its quiescent flux. We summarize in section 9.

\section{Energy-loss rates}

We first consider in detail the various processes
through which the pairs inside a new AGN jet component
lose energy. The blob is assumed to move
outward perpendicularly to the accretion disk plane with
velocity $c \, \beta_{\Gamma} = c \, \sqrt{1 - 1/\Gamma^2}$ 
where $\Gamma$ is the Lorentz factor of the bulk motion. 
The mechanisms that we take into account are inverse-Compton 
scattering of accretion disk photons, synchrotron and
synchrotron-self-Compton (SSC) losses. It is well-known that
energy exchange/loss due to elastic (M\o ller and Bhabha scattering) 
and inelastic scattering (pair bremsstrahlung emission) do
not contribute significantly for ultrarelativistic particles.
We consider them in detail in the second paper of this series,
dealing with mildly-relativistic pair plasmas. The same is true
for pair annihilation losses.

\subsection{External inverse-Compton losses}

We are now considering the single-particle energy loss rate due to
inverse-Compton scattering of external photons coming directly from a
central source, which we assume to be an accretion disk. We use the
accretion disk model of Shakura \& Sunyaev (1973) predicting, for
a central black hole of $M \sim 10^6$ -- $10^8 \, M_{\odot}$, a
blackbody spectrum according to a temperature distribution $T(R)$
given by

\vbox{
$$ \Theta(R) = {k_b \, T(R) \over m_e c^2} = 1.44 \, \left( {M \over 
M_{\odot}} \right)^{-{1 \over 2}} \, \cdot $$
\begin{equation}
\;\;\;\;\;\;
\cdot \, \left( {\dot M \over M_{\odot} / {\rm yr}} \right)^{1 \over 4} 
\, \left( {R \over R_g} \right)^{-{3 \over 4}} \, \left( 1 - \sqrt{ 
6 \, R_g \over R} \right)^{1 \over 4}. 
\end{equation}
}

In general, the energy loss rate due to inverse-Compton scattering
is given by the manifold integral

$$ - \left( {d\gamma \over dt} \right)_{IC} 
= {c \over 2} \int\limits_{-1}^1 
d\eta_e \> \int\limits_{-1}^1 d\eta_{ph} \int\limits_0^{2\pi} 
d\phi \int\limits_0^{\infty} d\epsilon \int\limits_{-1}^1 d\kappa' 
\int\limits_0^{2\pi} d\phi_s' $$
\begin{equation}
\int\limits_0^{\infty} d\epsilon_s' (1 - \beta \mu) 
{d\sigma_{KN} \over d\epsilon_s' \, d\Omega_{ph, s}'} 
\> n_{ph} (\epsilon, \Omega_{ph}) \, \Bigl( \gamma \, \epsilon_s' \, 
[1 + \beta \mu_s'] - \epsilon \Bigr) \end{equation}
where $\eta_e = \cos\theta_e$, $\eta_{ph} = \cos\theta_{ph}$, 
$\kappa' = \cos\chi'$,

\begin{figure}
\epsfxsize=6cm
\epsffile[0 250 420 550] {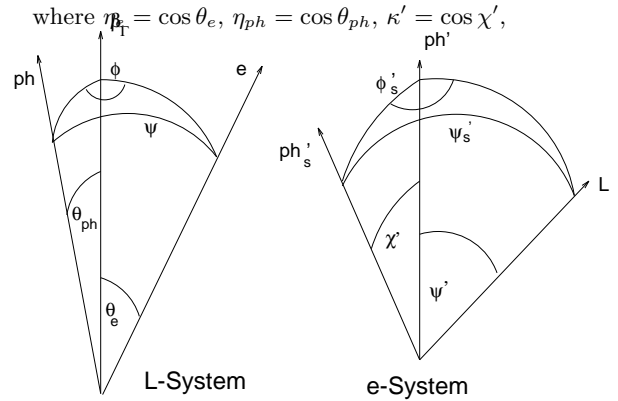}
\caption[]{Definition of the angles for calculation of the
inverse-Compton losses. 'L' in the right panel denotes the 
motion of the labor frame with respect to the electron
rest frame. The subscript `ph' refers to quantities of
the photon}
\end{figure}

\begin{equation}
\mu = \cos\psi = \eta_{ph} \, \eta_e + \sin\theta_{ph} \, \sin\theta_e 
\, \cos\phi
\end{equation}
\begin{equation}
\mu_s' = \cos\psi_s' = \kappa' \, \mu' + \sin\chi' \, \sin\psi' \, 
\cos\phi_s'.
\end{equation}
Here, primed quantities refer to the rest frame of the electron,
and the subscript 's' denotes quantities of the scattered photon.
The definition of the angles is illustrated in Fig. 2. Now, let the
superscript '$\ast$' denote quantities measured in the rest frame of
the accretion disk. Then, the differential number of accretion disk
photons in the blob frame is

$$ {d n_{ph} \over d\epsilon d\Omega_{ph}} = {\epsilon^2 \over 
\epsilon^{\ast \, 2}} \left({d n_{ph} \over d\epsilon d\Omega_{ph}} 
\right)^{\ast} = $$
\begin{equation}
{1 \over 2 \, c^3} \left( {m_e c^2 \over h} \right)^3 \, 
\epsilon^2 \left( e^{\epsilon {\Gamma (1 + \beta_{\Gamma} \eta_{ph}) 
\over \Theta (R)}} - 1 \right)^{-1} \> {1 + \beta_{\Gamma} \eta_{ph} 
\over \eta_{ph} + \beta_{\Gamma}}.
\end{equation}
Using this photon number and the full Klein-Nishina cross section, 
Eq. (5) becomes

$$ - \left( {d\gamma \over dt} \right)_{IC} = 
{\pi r_e^2 \, \Gamma^2 \over 4 \, z^2 \, c^2} 
\left( {m_e c^2 \over h} \right)^3 \int\limits_{-1}^1 d\eta_e 
\int\limits_{R_{\rm min}}^{R_{\rm max}} R \, dR \int\limits_0^{2\pi} d\phi 
$$
$$
\int\limits_0^{\infty} d\epsilon \> \epsilon^3 \left( e^{\epsilon 
{\Gamma (1 + \beta_{\Gamma} \eta_{ph}) \over \Theta(R)}} - 1 
\right)^{-1} \int\limits_{-1}^1 d\kappa \> (1 - \beta \mu) 
\left( \eta_{ph} + \beta_{\Gamma} \right)^2 \, \cdot
$$
\begin{equation}
{1 + F \, (F - 1) + F \, {\kappa'}^2 \over F^3} \left( \gamma^2 
{1 - \beta\mu + \beta \, \kappa' [\mu - \beta] \over F} - 1 \right)
\end{equation}
where

\begin{equation}
F = 1 + \epsilon \, \gamma (1 - \beta \mu) \, (1 - \kappa'),
\end{equation}
\begin{equation}
\eta_{ph} = {z - \beta_{\Gamma} \, \sqrt{R^2 + z^2} \over
\sqrt{R^2 + z^2} - \beta_{\Gamma} \, z}
\end{equation}
and $R_{\rm min}$ and $R_{\rm max}$ are the radius of the inner and 
outer edge of the accretion disk, respectively.

If all scattering occurs in the Thomson regime ($F \equiv 1$), we 
find (neglecting terms of order $1 / \gamma^2$):

$$ - \left( {d\gamma \over dt} \right)_{IC} \> \approx \> 
{4 \over 45} \gamma^2 {\pi^5 \, 
\sigma_T \over c^2} \, \Gamma^2 \, \left( {m_e c^2 \over h} 
\right)^3 \, \cdot $$
\begin{equation} \;\;\;\;\;\;\;\;\;\;
\int\limits_{ R_{\rm min}}^{R_{\rm max}} dR \, R \, \Theta^4 (R) \, 
{\left( x - \beta_{\Gamma} \, z\right)^2 \over  x^4}
\end{equation}
where $x = \sqrt{R^2 + z^2}$. Fig. 3 shows the 
energy loss rate computed using the full Klein-Nishina 
cross-section compared to the calculation in the 
Thomson limit as quoted above as well as the calculation
in the Thomson limit, combined with a point-source approximation
for the accretion disk (e. g. Dermer \& Schlickeiser 1993).

\begin{figure}
\rotate[r] {
\epsfxsize=6cm
\epsffile[100 20 600 50] {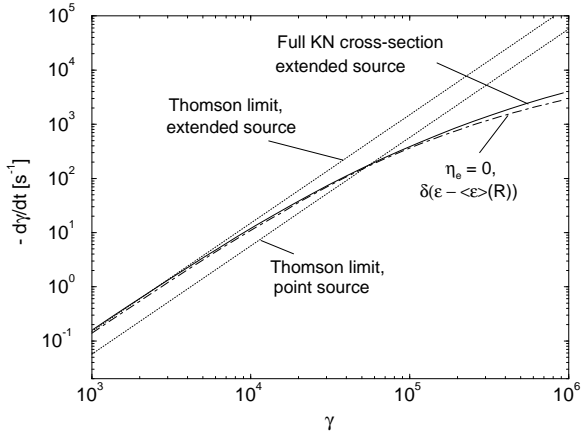} }
\caption[]{Energy loss rates of a test electron/positron due to
inverse-Compton scattering of accretion disk photons. Total disk
luminosity: $L_0 = 10^{44}$ erg s$^{-1}$, height: $z = 2 \cdot 10^{-4}$ pc}
\end{figure}

For electron energies of $\gamma \ll \Gamma / \langle \epsilon_{disk} 
\rangle$, Eq. (12) is a very good approach to the 
Inverse-Compton losses. In the case of small distances 
to the accretion disk ($z \ukl 10^{-2}$ pc in the case of
a disk luminosity $L_0 = 10^{46}$ erg s$^{-1}$; $z \ukl 10^{-3}$ pc 
for $L_0 = 10^{44}$ erg s$^{-1}$) the point-source approximation is 
not an appropriate choice.

A very useful approximation for all electron energies is based on
replacing the integration ${1 \over 2} \int_{-1}^1 d\eta_e$ by
setting $\eta_e = 0$. Furthermore, one can approximate the
thermal spectrum, emitted by each radius of the disk, by a
$\delta$ function in energy, $n(\epsilon, \Omega) \approx 
A \, \delta (\epsilon - <\epsilon> [R])$ where

$$ <\epsilon> (R) = {\Gamma (4) \zeta(4) \over
\Gamma(3) \zeta(3)} {\Theta(R) \over \Gamma 
(1 + \bg \eta_{ph})}. $$
With these simplifications, and neglecting terms of order 
$1/\gamma^2 \ll 1$, Eq. (9) becomes

$$ - \left({d\gamma \over dt}\right)_{IC} = \gamma^2 {\pi^5 \st 
\, \Gamma^2 \over 40 \, c^2} \left( {m_e c^2 \over h} \right)^3 
\int\limits_{R_{min}}^{R_{max}} R \, dR \, \cdot
$$
\begin{equation}
\;\;\;\;
\cdot \, \Theta(R)^4 {R \, (x - \bg z)^2 \over x^4} \, 
I (<\epsilon>, \gamma, \eta_{ph}) 
\end{equation}
where

$$
I(\epsilon, \gamma, \eta_{ph}) := \int\limits_0^{2\pi} d\phi
\> (1 - \beta\mu)^2 \, \cdot $$
\begin{equation}
\;\;\;\;\;\; \cdot \, \int\limits_{-1}^1 d\kappa' \> (1 - \kappa') \,
{1 + F^2 + F \, (\kappa'^2 - 1) \over F^4}
\end{equation}
and
$$ \mu = \sqrt{1 - \eta_{ph}^2} \, \cos\phi $$
$$ F =  1 + E \, (1 - a \, \cos\phi) \, (1 - \kappa') $$
$$ E = \epsilon\gamma, $$
$$ a = \beta \sqrt{1 - \eta_{ph}^2}. $$
The integral $I$ can be solved analytically, yielding

$$ I(\epsilon, \gamma, \eph) = {2 \, \pi \over E^2} \Biggl\lbrace
{6 \over E \, \sqrt{1 - a^2}} - {5 \over 6} + \ln\left( {1 + 2 \, E + N 
\over 2} \right) $$
$$ - {2 \over E} \, {\ln \left( {E \, a \over A} \right) \over 
\sqrt{1 - a^2}} - {8 \, D \over E \, a} \, {\ln( 1 - A \, D) \over 
1 - D^2} - {3 \over E^2} \, { \ln \left( {E \, a \over A} \right) 
\over (1 - a^2)^{3/2}} $$
$$ - {24 \, D^2 \over E^2 \, a^2 \, (1 - D^2)^2} \left({1 + D^2 
\over 1 - D^2} \, \ln [1 - A \, D] - {A \, D \over 1 - A \, D} \right) 
$$
\begin{equation}
\;\;\;\;\; + {2 \, (1 + 2 \, E)^2 + (2 \, E \, a)^2 \over 6 \> N^5} - 
{1 + 2 \, E \over 2 \, N^3} - {1 \over N} \Biggr\rbrace
\end{equation}
where

$$ N = \sqrt{ (1 + 2 \, E)^2 - (2 \, E \, a)^2}, $$
$$ A = {1 + 2 \, E \over 2 \, E \, a} - \sqrt{\left( {1 + 2 \, E 
\over 2 \, E \, a} \right)^2 - 1} \> < \> 1, $$
$$ D = {1 \over a} - \sqrt{ {1 \over a^2} - 1} \> < \> 1. $$
For small values of $E$, one should use the Taylor expansion of
Eq. (15), namely

$$ I(\epsilon, \gamma, \eph) \Bigr\vert_{E < {1 \over 2}} = 
2 \, \pi \, \Biggr\lbrace {8 \over 3} \, \left( 1 + {a^2 \over 2} 
\right) $$
\begin{equation}
- {56 \over 5} \, E \, \left( 1 + {3 \over 2} \, a^2 \right) + 
{92 \over 5} \, E^2 \, \left( 1 + 3 \, a^2 + a^4 \right) \Biggr\rbrace 
+ o \Bigl( E^3 \Bigr).
\end{equation}
This result (using Eq. [16] for $E < 0.05$) is illustrated by the 
dot-dashed curve in Fig. 3. In Fig. 5, the inverse-Compton losses 
(according to Eq. [13]) for set of parameters which is assumed to
be typical for BL Lac objects ($L_0 = 10^{44} {\rm erg \, s}^{-1}$, 
$z = 2 \cdot 10^{-4}$ pc, $\Gamma = 10$) are compared to the 
synchrotron and SSC energy losses derived in the following subsections.

\subsection{Synchrotron and synchrotron-self-Compton losses}

If, for the synchrotron (sy) losses, we neglect inhomogeneities and 
effects of anisotropy, we have

$$
\left( {d \gamma \over dt} \right)_{sy} = - {4 \over 3} c \, \sigma_T
{B^2 \over 8 \pi m_e c^2} (\gamma_t^2 - 1) $$
\begin{equation}
\hskip 2.5cm
\approx - \, 3.25 \cdot 10^{-8} \; B_0^2 \> \gt^2 \; {\rm s}^{-1}
\end{equation}
(e. g., Rybicki \& Lightmann 1979) where $B_0 := B / (1 \, {\rm G})$. 

Evaluating the single-particle energy loss rate due to the SSC process, 
we restrict ourselves to regarding only single scattering events. If 
the particle distribution functions are given by power-laws, the 
synchrotron photon spectrum is also described by a power-law. 
However, since we are intrested in the detailed shape of the
distributions, deviating from a simple power-law, we have to
calculate the synchrotron spectrum in more detail.

In an optically thin source, the differential number of synchrotron 
photons in the energy interval $[\epsilon, \epsilon + d\epsilon]$ is 
given by

\vbox{
$$ n_{sy} (\epsilon) = $$
\begin{equation}
\;\;\;\;\;\;\;
{R_B \over 3 \, \, c} \, {m_e c^2 \over h} \sum\limits_{\pm}
\int\limits_1^{\infty} d\gpm \> { P_{\nu} (\gpm) \over h \, \nu } 
\Biggr\vert_{\nu = \epsilon \, {m_e c^2 \over h}} \> n_{\pm} (\gpm).
\end{equation} 
}
The summation symbol denotes the sum of the contributions from
electrons and positrons to the synchrotron spectrum. Averaging over
all pitch-angles of the electrons, the spectral emissivity 
$P_{\nu} (\gamma)$ of a single electron of Lorentz factor 
$\gamma$ can be expressed as

\begin{equation}
{P_{\nu} (\gamma) \over h \, \nu} = {\alpha \, \pi \over 
\sqrt 3 \, \gamma^2} \, \left( 1 + \left[ \gamma {\nu_p \over \nu}
\right]^2 \right) \, CS (x)
\end{equation}
(Crusius \& Schlickeiser 1986) where $\alpha$ is the finestructure constant,

\begin{equation}
x = {2 \, \nu \over 3 \, \nu_e \, \gamma^2} \, \left( 1 + \left[ \gamma
{\nu_p \over \nu} \right]^2 \right)^{3 \over 2},
\end{equation}
$\nu_e$ is the non-relativistic electron/positron gyrofrequency, $\nu_p$
is the plasma frequency of the relativistic pair plasma, i. e.

\begin{equation}
\nu_p = \sqrt{ n_e \, e^2 \over \pi \, m_e \, \langle \gamma \rangle}
\end{equation}

\begin{equation}
CS(x) = W_{0, {4 \over 3}} (x) \, W_{0, {1 \over 3}} (x) \> - 
\> W_{{1 \over 2}, {5 \over 6}} (x) \, W_{-{1 \over 2}, {5 \over 6}} (x)
\end{equation}
and $W_{\kappa, \mu}$ are the Whittaker functions.

If the particle density exceeds $n \ugr 10^8$ cm$^{-3}$ synchrotron 
self absorption (SSA) is negligible for the following two reasons:
The optical depth due to synchrotron self absorption for an initial
power-law distribution $n(\gamma) = n_0 \, \gamma^{-2}$ can be 
estimated as

\begin{equation}
\tau_{\rm SSA} \approx {\sqrt 3 \, \alpha \, \hbar \, c^2 \over
m_e c^2 } \, \Gamma\left( {5 \over 3} \right) \, \Gamma\left( 
{4 \over 3} \right) \, n_0 \, \nu_e^2 \, \nu^{-3} \, R_B
\end{equation}
(Schlickeiser \& Crusius 1989). This optical depth becomes unity for 

\begin{equation}
\nu_{\rm SSA} = 5.7 \cdot 10^{10} \> \gamma_{1;3}^{1/3} \, B_{-1}^{2/3} 
\, n_9^{1/3} \, R_{11}^{1/3} \; {\rm Hz}
\end{equation}
where $\gamma_{1;3}$ is the lower cutoff Lorentz factor in units of
$10^3$, $B_{-1}$ is the magnetic field in units of $0.1$ G, $n_9$ is
the pair density in units of $10^9$ cm$^{-3}$ and $R_{11}$ is the
blob radius in units of $10^{11}$ cm. The frequency $\nu_{\rm SSA}$ is 
of the same order as the Razin-Tsytovich frequency

\begin{equation}
\nu_R = 2 \cdot 10^{11} \> n_9 \, B_{-1}^{-1} \, 
\langle\gamma\rangle^{-1/2} \; {\rm s}^{-1}
\end{equation}
where the luminosity is strongly suppressed due to plasma effects.
Here, $\langle\gamma\rangle$ denotes the average electron Lorentz
factor. This demonstrates that for pair densities $n \ugr 10^8$ cm$^{-3}$
synchrotron self absorption can be neglected. For lower densities
we include it in our calculations, evaluating

\begin{equation}
\tau_{\rm SSA} = - R_B {c^2 \over 8 \pi \nu^2 m_e c^2}
\int\limits_1^{\infty} d\gamma \> P_{\nu} (\gamma) \, \gamma^2
{d \over d\gamma} \left( {n[\gamma] \over \gamma^2} \right)
\end{equation}
(e. g., Rybicki \& Lightman, 1979) self-consistently.

From the point of view of reacceleration, synchrotron self absorption
can be neglected since particles resonating with photons at the lower 
cut-off of the synchrotron spectrum should have Lorentz factors of

\begin{equation}
\gamma_{res} \approx 450 \> \gamma_{1;3}^{1/6} \, B_{-1}^{-1/6} \,
n_9^{1/6} \, R_{11}^{1/6}
\end{equation}
which is lower than the particle Lorentz factors we deal with in the 
simulations carried out in this paper. 

If all scattering occurs in the Thomson limit, $\gpm \, \epsilon \ll 1$, 
the first order SSC energy loss rate is easily determined by

\begin{equation}
\left( {d\gt \over dt} \right)_{\rm SSC} = - {4 \over 3} c \, \sigma_T \,
\left( \gpm^2 - 1 \right) \int\limits_0^{\infty} d\epsilon \> \epsilon
\> n_{sy} (\epsilon).
\end{equation}
where the synchrotron luminosity can be calculated as

\begin{equation}
\int\limits_0^{\infty} d\epsilon \> \epsilon \, n_{sy} (\epsilon) = 
{R_B \over 3 \, c} \int\limits_1^{\infty} d\gpm n_{\pm} (\gpm) 
\, \left\vert \left( {d\gpm \over dt} \right)_{sy} \right\vert, 
\end{equation}
yielding for the initial power-law distribution functions

$$
\left( {d\gt \over dt} \right)_{\rm SSC} \approx 
$$
\begin{equation}
{16 \over 27} c \, \st^2 \left( {B^2 \over 8 \pi \, m_e c^2} \right) 
R_B \, n_{\pm} \, \gzpm^{3 - s} \gepm^{s - 1} \, {s - 1 \over 3 - s} 
\> \gamma^2.
\end{equation}
Discarding the Thomson approximation and using the notation of the 
previous subsection, the exact expression for the SSC energy losses 
is

$$ - \left( {d\gamma \over dt} \right)_{\rm SSC} = {c \, \pi \, r_e^2 
\over 2} \int\limits_{-1}^1 d\mu \int\limits_{-1}^1 d\kappa' 
\int\limits_0^{\infty} d\epsilon \> \epsilon \, n_{ph} (\epsilon)
\, (1 - \beta \mu) \, \cdot $$
\begin{equation}
\cdot \, {1 + F \, (F - 1) + F \, {\kappa'}^2 \over
F^3} \left( \gamma^2 { 1 - \beta\mu + \beta\kappa' \, [\mu - \beta]
\over F} - 1 \right).
\end{equation}

\begin{figure}
\rotate[r] {
\epsfxsize=6cm
\epsffile[100 20 600 50] {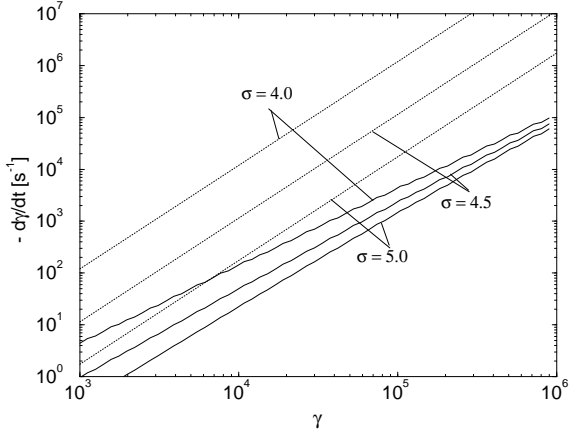} }
\caption[]{Energy loss rates of a test electron/positron due to
inverse-Compton scattering of synchrotron photons. $B = 1$ G, $R_B
= 10^{15}$ cm, $n_e = 10^4$ cm$^{-3}$. Thomson-scattering (dotted
curves) and exact Klein-Nishina cross-section (solid curves)}
\end{figure}

In this case of an isotropic radiation field in the blob frame, 
the integrations over $\mu$ and $\kappa'$ can be solved analytically 
if we neglect terms of order $1 / \gamma^2 \ll 1$ and of order 
$\epsilon / \gamma \ll 1$. This yields

\begin{equation}
- \left( {d\gamma \over dt} \right)_{\rm SSC} = c \, \pi \, r_e^2 
\int\limits_0^{\infty} d\epsilon \> n_{ph} (\epsilon) \> 
{G(\gamma\epsilon) \over \epsilon},
\end{equation}
where $c \, \pi \, r_e^2 \, G(\gamma\epsilon) / \epsilon$ cm$^{-3}$ 
is the energy loss rate of an electron of energy $\gamma$ scattering
an isotropic, monochromatic radiation field of photon energy 
$\epsilon$ and photon density 1 cm$^{-3}$ and

\vbox{
$$ G(E) = {8 \over 3} \, E \, {1 + 5 \, E \over (1 + 4 \, E)^2} 
- {4 \, E \over 1 + 4 \, E} \left( {2 \over 3} + {1 \over 2 \, E} + 
{1 \over 8 \, E^2} \right) $$
$$ + \ln (1 + 4 \, E) \left( 1 + {3 \over E} + {3 \over 4} \, 
{1 \over E^2} + {\ln[1 + 4 \, E ] \over 2 \, E} - {\ln [4 \, E] 
\over E} \right) $$
\begin{equation}
\;\;\;\;\;\;\;\;\;\;\;
- {5 \over 2} {1 \over E} + {1 \over E} \sum\limits_{n=1}^{\infty} 
{(1 + 4 \, E)^{-n} \over n^2} - {\pi^2 \over 6 \, E} - 2.
\end{equation}
}

For small values of $E = \gamma \epsilon$ (i. e. the Thomson
limit) the Taylor expansion

\begin{equation}
G(E) \biggr\vert_{E < {1 \over 4}} = E^2 \, \left\lbrace 
{32 \over 9} - {112 \over 5} E + {3136 \over 25} E^2 + o \bigl( E^3 
\bigr) \right\rbrace
\end{equation}
to lowest order reduces Eq. (32) to Eq. (28) for $\gamma^2 
\gg 1$. In Fig. 4, the analytic solution (32) is compared to the
Thomson scattering result (Eq. [27]).

\begin{figure}
\rotate[r] {
\epsfxsize=6cm
\epsffile[100 20 600 50] {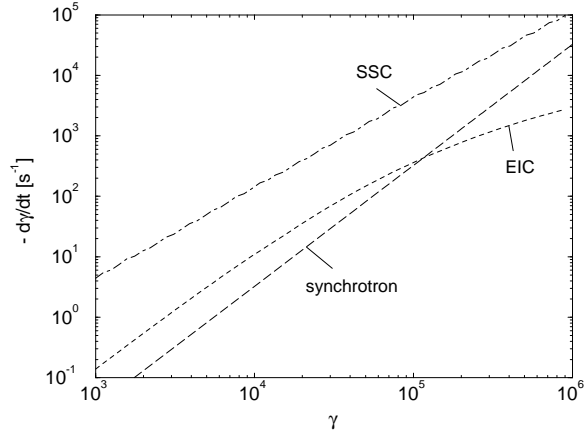} }
\caption[]{Energy loss rates of a test particle (Lorentz
factor $\gt$) due to inverse-Compton scattering of external
(accretion disk) photons (EIC; disk luminosity $L_0 = 10^{44} 
{\rm erg \, s}^{-1}$, $M_{BH} = 10^6 \, M_{\odot}$, $z = 2 \cdot 
10^{-4}$ pc, $\Gamma = 10$) and of synchrotron photons (SSC; 
$B = 1$ G, $R_B = 10^{15}$ cm, $n_{\pm} = 10^4$ cm$^{-3}$, 
$\sigma = 4.0$), and to synchrotron emission ($B = 1 \, G$)}
\end{figure}

It should be noted that for such an ultrarelativistic particle
distribution Compton scattering its own synchrotron photons,
Klein-Nishina corrections do not lead to a break in the energy
dependence of the energy-loss rate, but due to the fact that
with increasing energy a decreasing fraction of the scattering
events occurs in the Thomson regime, lead to a much smoother
flattening of the energy depencence than what is obtained from
the relatively sharp soft photon distribution coming from the
accretion disk. In the energy regime considered here, the
energy-loss rate scales as $- d\gamma/dt \, \sim \, \gamma^{\alpha}$
with $\alpha \approx 1.5$ for the spectral index $s \sim 2$ of the 
particle distributions $n(\gamma) \sim \gamma^{-s}$. Clearly,
this effect depends strongly on the temporal particle distribution
which determines the synchrotron spectrum.

The result of Eq. (32) (using Eq. [34] for $E < 0.1$) is 
included in Fig. 5 for a magnetic field strength of 
${\rm B} = \, 1$~G and a blob radius of $R_B = 10^{15}$ cm. 
Higher-order SSC scattering (up to $n$th order in the $n$th time step) 
is incorporated in our simulations by replacing $n_{sy} (\epsilon)$ by 
$n_{sy} (\epsilon) + n_{\rm SSC} (\epsilon)$ in Eq. (32).

\section{Discussion of the elementary processes}

From Fig. 5 we can deduce some estimates about which of the 
various energy loss and exchange processes discussed above play 
an important role in different energy regimes of particles.

First, we note that the energy transfer rates due to collisional 
effects (elastic scattering, bremsstrahlung emission) and the pair 
annihilation rate scale linearly with particle density $n$. The 
synchrotron and SSC losses basically scale as $\sim B^2$ (the SSC 
losses, additionally, $\sim R_B \cdot n$), whereas the external 
inverse-Compton losses basically scale as $L_0 \, / \, z^2$. The 
SSC losses also strongly depend on the steepness of the initial 
distribution functions since a harder particle spectrum yields a 
harder synchrotron spectrum which, in turn, leads to higher 
inverse-Compton losses.

The effect of elastic scattering is of minor importance for 
ultrarelativistic particles. We find the time-scale for 
thermalization to be given by

\begin{equation}
\tau_{\rm th} := {\gt - 1 \over \left\vert d\gt / dt \right\vert_{\rm
elast. \, scat.}} \approx {2.5 \cdot 10^{12} \over n_4} \> {\rm s}
\end{equation}
for $\gt = 100$ which increases rapidly towards higher particle 
energies. Here, $n_4$ is the particle density in units of 
$10^4$ cm$^{-3}$.

The distribution functions' high-energy tail will be most heavily 
influenced by one of the radiative energy-loss processes discussed 
above. Of course, the relative importance of the different radiation 
processes is strongly dependent on the exact values and variation 
of parameters, especially the magnetic field strength and the 
luminosity of the central photon source.

The time-scales for the respective processes can be estimated as

\begin{equation}
\tau_{\rm SSC} := {\gt - 1 \over \left\vert d\gt / dt \right\vert_{\rm SSC}}
\approx {5 \cdot 10^{10} \over \gt^{1/2} \, n_4 \, B_0^2 \, R_{11}} 
\> {\rm s}
\end{equation}
--- where $R_{11} = {R_B \over 10^{11} \, {\rm cm}}$ --- for the 
synchrotron-self-Compton process,

\begin{equation}
\tau_{sy} \approx {3 \cdot 10^7 \over \gt \, B_0^2} \; {\rm s}.
\end{equation}
for the synchrotron process,

\begin{equation}
\tau_{EIC} \approx {10^5 \over \gt} {\Gamma_1^2 \, z_{-3}^2 \over L_{46}} 
\; {\rm s}
\end{equation}
where $\Gamma_1 = {\Gamma \over 10}$, $L_{46} = {L \over 10^{46} {\rm erg 
\, s^{-1}}}$ and $z_{-3} = {z \over 10^{-3} \, {\rm pc}}$ for the 
external inverse-Compton losses in the Thomson regime, and

\begin{equation}
\tau_{\rm bremsstr.} \approx {3.2 \cdot 10^{11} \over n_4} \> \gt^{-0.1}
\> {\rm s}.
\end{equation}

In the beginning phase of the jet evolution, the effect of pair
annihilation makes a negligible contribution to the
distribution functions' evolution. The annihilation timescale for
particles near the lower cutoff can be estimated as

\begin{equation}
\tau_{pa} := {n(\gt) \over \left\vert dn(\gt) / dt \right\vert_{pa}}
\approx {10^{12} \over n_4} \, \gt \> {\rm s}
\end{equation}
which indicates that pair annihilation and bremsstrahlung are of 
minor importance to ultrarelativistic pair plasmas of density 
$\sim 10^9$ cm$^{-3}$.

These considerations and the investigation of the resulting
photon spectra lead us to some general simplifications in
the treatment of the particle spectra evolution as long 
as the plasma is ultrarelativistic: 

\noindent
(a) Pair annihilation is negligible.

\noindent
(b) Elastic (M\o ller and Bhabha scattering) and inelastic 
scattering (bremsstrahlung emission) are only important for 
relatively low-energetic particles ($\gamma < 100$) which do 
not exist in our model system during the phase we are dealing 
with in this paper.

\noindent
(c) The effect of energy dispersion is neglected.

\noindent
(d) Bremsstrahlung emission is negligible for pair densities
$n \ukl 10^7$ cm$^{-3}$.

\section{The $\gamma$-ray photosphere}

We now discuss the initial height of a plasma blob from
which TeV $\gamma$-ray emission as observed from Mrk~421
can escape freely. Since BL Lac objects have very dilute emission
line clouds, backscattering of accretion disk radiation
into the trajectery of emitted $\gamma$-rays is negligible.

\begin{figure}
\rotate[r] {
\epsfxsize=6cm
\epsffile[100 20 600 50] {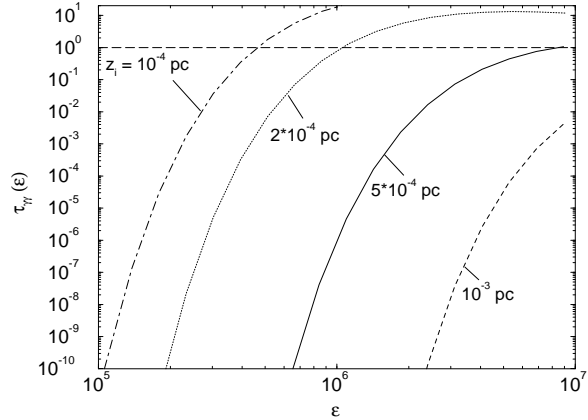} }
\caption[]{Pair production optical depth for high-energy $\gamma$-rays,
emitted at different heights ($z_i$), interacting with direct accretion 
disk radiation ($L = 10^{44}$ erg/s, $M = 10^6 \, M_{\odot}$) }
\end{figure}

Then, the optical depth for a photon of dimensionless energy
$\epsilon_1$, emitted at height $z_i$ due to interaction with 
accretion disk radiation is calculated as

$$ \tau_{\gamma\gamma} (\epsilon_1, z_i) = 2 \, \pi \, \cdot $$
\begin{equation} 
\cdot \int\limits_{z_i}^{\infty} dz \int\limits_{-1}^1 d\mu \, 
(1 - \mu) \!\!\! \int\limits_{2 \over \epsilon_1 \, (1 - \mu)}^{\infty} 
\!\!\! d\epsilon \> \sigma_{\gamma\gamma} (\epsilon_1, \epsilon, \mu) 
\, n_{ph} (\epsilon, \Omega; z)
\end{equation}
where $n_{ph} (\epsilon, \Omega; z)$ is the differential number of 
accretion disk photons at height $z$, $\sigma_{\gamma\gamma}$ is
the pair production cross section 

$$ \sigma_{\gamma\gamma} (\beta) = $$
\begin{equation}
{3 \over 16} \sigma_T \, (1 - \beta^2) \, \left( \left[ 3 - 
\beta^4 \right] \ln\left[ {1 + \beta \over 1 - \beta} \right] - 
2 \, \beta \, \left[ 2 - \beta^2 \right] \right), 
\end{equation}
\begin{equation}
\beta = \sqrt{1 - {2 \over \epsilon \epsilon_1 \, (1 - \mu)}}
\end{equation}
and we assumed for simplicity that the $\gamma$-ray is emitted 
along the symmetry axis. The result for different injection
heights $z_i$ is illustrated in Fig. 6. and shows that the
$\gamma$-ray photosphere for $\gamma$-rays of energy 500 GeV
-- 1 TeV is located slightly above $2 \cdot 10^{-4}$ pc in 
the case of an accretion disk of luminosity $L = 10^{44}$ 
erg s$^{-1}$.

\section{Simulations of the particle evolution}

Now that we know the energy-loss rates of the pairs inside
a new jet component, we can follow the evolution of such a
pair plasma. In the first phase in which radiative losses
are the only really important mechanisms, this can be done
most easily by following the cooling of single particles and 
using the fact that the total particle number is conserved:
Let

\begin{equation}
f(\gamma [t], t) = {d\gamma \over dt} \end{equation}
be the cooling rate for one particle. From $t_0$ to 
$t = t_0 + \Delta t$ the particle energy reduces from
$\gamma_0$ to $\gt$ where

\begin{equation}
\gt \approx \gamma_0 + \Delta t \> f(\gamma_0, t_0).
\end{equation}

Particle number conservation implies

\begin{equation}
n_t (\gt) = n_0 (\gamma_0) \, {d\gamma_0 \over d\gamma_t}.
\end{equation}
If we choose $\Delta t$ small enough so that the explicit
time dependence of $f$ (which, e. g., represents the temporal 
variation of the background photon distributions) within one 
time interval is negligible, we find

\begin{equation}
{d\gamma_0 \over d\gt} \approx {f(\gamma_0, t_0) \over 
f(\gt, t_0)}.
\end{equation}
This is used to follow the cooling of the injected pair
plasma. In each time step, we calculate the emanating photon
spectra (see Sect. 6) and use the respective synchrotron and
SSC spectra in order to calculate the SSC cooling-rate
according to Eq. (32) self-consistently.

\section{Photon spectra}

From the observational point of view, it is most interesting
to know both the evolution of the radiation spectra with
time and the time-integrated photon spectra, which are emanating 
from the pair plasmas treated in the previous sections. We first 
evaluate all the contributions to the photon spectra in the frame of 
reference of the jet (blob frame) and then transform to the observer's 
frame using

\begin{equation}
\dot N^{\ast}_{ph} \left( \epsilon^{\ast}, \Omega^{\ast}, t^{\ast} 
\right) = D^2 \dot N_{ph} \left( { \epsilon^{\ast} \over D} , \Omega, 
t \right)
\end{equation}
where the asterisk denotes quantities measured in the observer's
frame (which we idenfity with the accretion-disk frame), 
$\dot N(\epsilon, \Omega) = V_B \, \dot n(\epsilon, \Omega)$ 
is the spectral photon production rate in the energy interval 
$[\epsilon, \, \epsilon + d\epsilon]$ ($\epsilon = h \nu / 
(m_e c^2)$) in the solid angle interval
$[\Omega, \, \Omega + d\Omega]$, integrated over the whole blob 
volume $V_B$, $D = \left( \Gamma [1 - \beta_{\Gamma} \mu^{\ast}] 
\right)^{-1}$ is the Doppler factor, $\mu^{\ast} = \cos\theta^{\ast}$ 
is the cosine of the observing angle in the observer's frame and 
$\mu = {\mu^{\ast} - \beta \over 1 - \beta \mu^{\ast}}$ is the 
observing angle cosine in the blob frame. A time interval $\Delta 
\, t$ measured in the blob frame is related to a light reception 
time interval $\Delta \, t_r$ by $\Delta \, t_r = \Delta \, t \, / 
\, D$ and to an accretion disk frame time interval $\Delta \, 
t^{\ast}$ by $\Delta \, t^{\ast} = \Gamma \, \Delta \, t$.

The most important contributions to the $\gamma$-ray spectra
in the MeV -- TeV regime come from inverse-Compton interactions.
The photon spectrum resulting from inverse-Compton scattering of
accretion disk photons is calculated as

$$
\dot n_{EIC} (\epsilon_s, \Omega_s) = c \int\limits_0^{\infty} d\epsilon
\int\limits_{4\pi} d\Omega_{\epsilon} \> n_{ph} (\epsilon, 
\Omega_{\epsilon}) \, \cdot 
$$
\begin{equation}
\;\;\;\;\;\;\;\;\;\;\;\;\; \cdot \int\limits_1^{\infty} d\gamma
\int\limits_{4\pi} d\Omega_{e} \> {n_e (\gamma) \over 4 \, \pi} 
\,  (1 - \beta \mu) \, {d^2 \sigma_C \over d \epsilon_s \, d\Omega_s} 
\end{equation}
where $\epsilon_s = h \nu_s / (m_e c^2)$ is the normalized energy 
and $\Omega_s$ the solid angle of the motion of the scattered photons, 
$n_{ph} (\epsilon, \Omega_{\epsilon})$ is the differential number 
density of accretion disk photons at the location of the scattering 
event as discussed in Section 2.1 and the angle variables are 
illustrated in Fig. 2 and Eqs. (6) and (7). 

It has turned out to be 
most convenient for this calculation (using the full Klein-Nishina 
cross section) to use the differential Compton scattering cross 
section in the blob frame:

\vbox{
$$
{d^2 \sigma_C \over d \epsilon_s \, d\Omega_s} = {1\over 2} { r_e^2 \over
\gamma^2 \, (1 - \beta \mu)^2 } \left( {\epsilon_s \over 
\epsilon}\right)^2
\cdot
$$
$$ \cdot \delta \left[ \epsilon_s - \epsilon { \gamma \, (1 - \beta \mu)
\over \epsilon (1 - \kappa) + \gamma \, (1 - \beta \mu_s) }\right]
\cdot $$
\begin{equation}
\;\;\;\;\;\;\;\;\;\;\;\;\;
\cdot \Biggl( {\epsilon [1 - \beta \mu] \over \epsilon_s [1 - \beta 
\mu_s]} + {\epsilon_s [1 - \beta \mu_s] \over \epsilon [1 - \beta \mu]} 
- 1 + {\kappa'}^2 \Biggr)
\end{equation} }
where
\vbox{
$$
\kappa' = 1 - {1 - \kappa \over \gamma^2 \, (1 - \beta\mu) \,
(1 - \beta\mu_s)} $$
\begin{equation}
\hskip 1.5cm = \; 1 + {1 \over \gamma \epsilon (1 - \beta \mu) } -
{1 \over \gamma \epsilon_s (1 - \beta \mu_s) }
\end{equation} }
(Jauch \& Rohrlich 1976). (Of course, using the KN cross section 
in the electron system and transforming the final photon state 
back to the blob frame leads to the same final expression for the
spectrum of scattered photons. It is more straightforward to 
use directly the cross section [50].)

With the assumptions we made above, the SSC spectrum is isotropic
in the blob frame, and

$$
{\dot n_{\rm SSC} (\epsilon_s) \over 4 \, \pi} = c \int\limits_0^{\infty} 
d \epsilon \> {n_{sy} (\epsilon) + n_{\rm SSC} (\epsilon) \over 4 \, \pi} 
\int\limits_{4\pi} d\Omega_{\epsilon} \> \cdot
$$
\begin{equation}
\;\;\;\;\;\;\;\;\;\;\;\;\;\;\;
\int\limits_1^{\infty} d\gamma \> {n_e (\gamma) \over 4 \, \pi} 
\int\limits_{4\pi} d\Omega_e \> (1 - \beta \mu) \, {d^2 \sigma_C 
\over d \epsilon_s \, d\Omega_s}.
\end{equation}
Technically, we include higher order SSC scattering by using in each 
time step the SSC photon density $n_{\rm SSC}$ of the foregoing time 
step together with the synchrotron photon density as seed photon field
for SSC scattering. 

Using the cross section (50), Eq. (52) can be rewritten as

$$ 
\dot n_{\rm SSC} (\epsilon_s) = {c \, r_e^2 \over 4} \int\limits_1^{\infty}
d\gamma \> {n_e (\gamma) \over \gamma^2} \int\limits_{-1}^1 d\kappa 
\int\limits_{-1}^1 d\mu_s \int\limits_0^{2\pi} d\rho \, \cdot $$
\begin{equation}
\;\;\;\; \cdot \, { n_{ph} (F \, \epsilon_s) \over 1 - \beta \, \mu_s} 
\Biggl\lbrace {\gamma \, (1 - \beta \, \mu) \over N} + {N \over \gamma 
\, (1 - \beta \, \mu) } - 1 + {\kappa'}^2 \Biggr\rbrace
\end{equation}
where $\rho$ is the azimuthal angle between photon and electron
direction of motion around the direction of motion of the scattered
photon, the other angle variables are the same as in Section 2.1,
$\kappa'$ is given by Eq. (52), $n_{ph}$ denotes the sum of
synchrotron and previously produced SSC photons, and

\begin{equation}
 N := \gamma \, (1 - \beta \, \mu) - \epsilon_s \, (1 - \kappa),
\end{equation}
\begin{equation}
F = \gamma \, {1 - \beta \, \mu_s \over N},
\end{equation}
\begin{equation}
\mu = \kappa \, \mu_s + \sqrt{1 - \kappa^2} \, \sqrt{1 - \mu_s^2} \, 
\cos\rho.
\end{equation}

\begin{figure*}
\rotate[r] {
\epsfxsize=12cm
\epsffile[20 0 600 450] {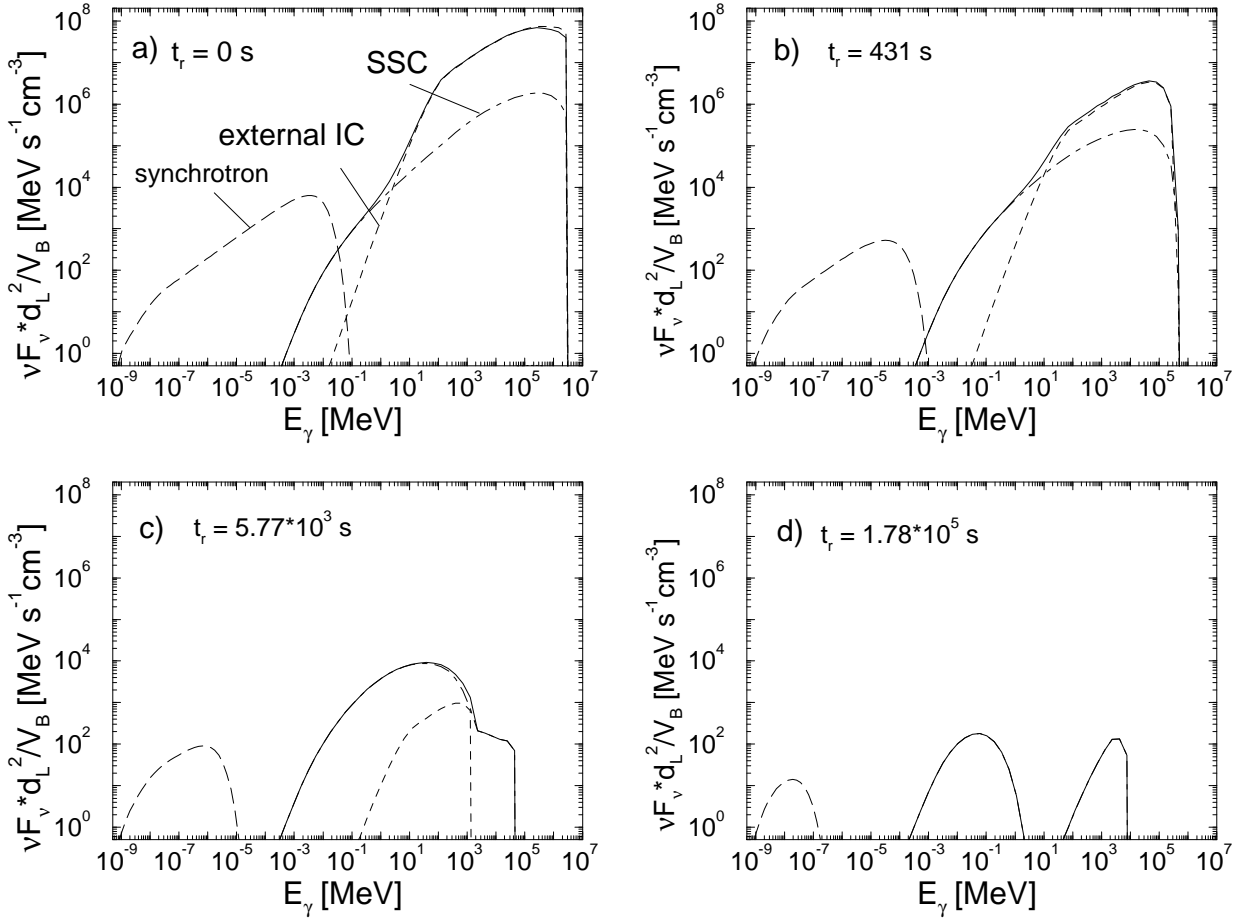} }
\caption[]{Instantaneous broadband spectra from the blob
of simulation 1. Parameters: $\gamma_1 = 10^3$, $\gamma_2 = 10^6$,
$s = 2.0$, $R_B = 5 \cdot 10^{14}$ cm, $n = 10^4$ cm$^{-3}$, 
$B = 0.05$ G, $L_{\rm disk} = 10^{44} \, {\rm erg \, s}^{-1}$, 
$M_{BH} = 10^6 \, M_{\odot}$ $z_i = 5 \cdot 10^{-4}$ pc, 
$\Gamma = 20$, $\theta = 5^0$.
Total emission (corrected for $\gamma$-$\gamma$ absorption): 
solid; IC scattering of accretion disk 
radiation: dashed; SSC radiation: dot-dashed; synchrotron 
radiation: long dashed. (all quantities in the observer's frame) }
\end{figure*}

If pair bremsstrahlung can contribute significantly to the time-integrated
photon spectra (which we find to be the case if $n \ugr 10^6$ cm$^{-3}$) 
we evaluate the rate of production of bremsstrahlung photons per unit
volume, $\dot n_{br}$, in the comoving fluid frame as

$$
\dot n_{br} (\epsilon) =  4 \, c \int d\gm n_- (\gm) \int d\gp n_+ (\gp) 
\cdot
$$
\begin{equation}
\;\;\;\;\;\;\;\;\;\;\;
\cdot \int d\gcm {\bcm \gcm^3 \over \bm \gm^2 \, \bp \gp^2} {d\sigma
\over d \epsilon} (\epsilon, \gm, \gp, \gcm)
\end{equation}
where we use the differential cross section

$$ \left( {d\sigma \over d\epsilon} \right)_{cm} = {8 \, r_0^2 \, 
\alpha \over k}  {\gcm - \ecm \over \gcm} \left( \ln 
\left[ 4 \, \gcm^2 {\gcm - \ecm \over \ecm} \right] - 
{1 \over 2} \right) \cdot $$
\begin{equation}
\cdot \left( {\gcm - \ecm \over \gcm} + {\gcm \over \gcm - \ecm} -
{2 \over 3} \right)
\end{equation}
(Alexanian 1968) in the center-of-momentum frame of the scattering
particles, $\alpha$ is the hyperfinestructure constant, and the 
integration limits are given by

\begin{equation}
\;\;\;\;\; \epsilon < \gpm \bpm^2 \;\;\;\;\;\;\; {\rm and} \;\;\;\;\;\;\;
\gcm^2 \ge {\gpm - {\epsilon \over 4} \over \gpm - \epsilon}.
\end{equation}

The calculation of the synchrotron spectrum has been described in
section 2.2. Pair annihilation radiation is negligible for an
ultrarelativistic pair plasma.

After having calculated the resulting synchrotron and $\gamma$-ray
spectra, we have to check whether $\gamma$-rays of energies
$\sim 1$ TeV can escape the emitting region. For this purpose
we calculate the optical depth due to $\gamma$-$\gamma$ pair 
production of high-energy $\gamma$-rays interacting with the
radiation produced in the blob. This calculation is carried out
in the blob frame where we assume all contributions
to be isotropic (which is not the case for the external IC
component; but for a $\gamma$-ray traveling with small angle
with respect to the jet axis, the assumption of isotropy even
overestimates the pair production optical depth). The results
show that even in the first time step (which, of course, is 
the most critical one) the $\gamma$-$\gamma$ optical depth
does not exceed one for parameters suitable to fit the broadband 
spectrum of Mrk~421 (see section 8). Nevertheless, we include the 
small effect of $\gamma$-$\gamma$ absorption when calculating the 
emanating photon spectra which is the reason for the 
total emission (corrected for $\gamma$-$\gamma$ absorption) 
being slightly lower than the EIC component in Fig. 7 a. 
The injection of pairs due to $\gamma$-$\gamma$ pair 
production (for details see B\"ottcher \& Schlickeiser
1997) is negligible even in the case of relatively
high density as assumed for Figs. 7 and 8.

\section{General Results}

In this section we present some general results which
we collected during a series of simulations with varying
parameters. 

An interesting feature is that the time-averaged
SSC emission of a pair distribution with a lower 
cut-off $\gamma_1 \ugr 100$ follows a power-law
in the X-ray range with energy spectral index $0.5 \ukl 
\alpha \ukl 0.7$ which is consistent with the X-ray spectra
of flat-spectrum radio quasars (FSRQs) and low-frequency
peaked BL Lacs (LBLs) where the hard X-ray emission
can be attributed to jet emission. The X-ray spectrum
slightly hardens with increasing lower cut-off of
the particle injection spectrum.

\begin{figure}
\rotate[r] {
\epsfxsize=5.8cm
\epsffile[100 20 600 50] {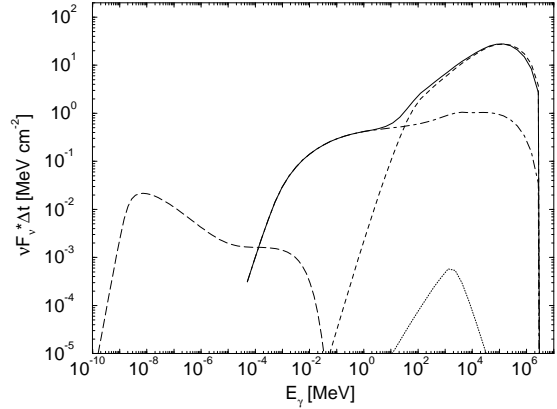} }
\caption[]{Time-integrated broadband spectrum from the evolving
blob illustrated in Fig. 7 (total $\gamma$-ray spectrum: solid; 
EIC: dashed; SSC: dot-dashed; bremsstrahlung: dotted; synchrotron: 
long dashed) }
\end{figure}

Our simulations showed that assuming a particle 
spectrum extending up to $\gamma_2 \sim 10^5$ -- 
$10^6$ where some EIC scattering occurs in the 
Klein-Nishina regime, the EIC spectrum is harder 
than in the case of pure Thomson scattering where 
we recover the classical result $\dot n (\epsilon) 
\sim \epsilon^{-(s+2)/2}$ (Dermer \& Schlickeiser 
1993 a) above the break in the photon spectrum 
caused by incomplete Compton cooling. 

A harder time-integrated EIC spectrum (assuming the
same spectral index of the injected pair distributions)
can also result if cooling due to EIC is very inefficient
(i. e. for a blob at large distance from the accretion
disk) implying that the effect of dilution of the
soft photon field according to its $z^{-2}$ dependence
dominates the evolution of the EIC spectrum.

\begin{figure*}
\rotate[r] {
\epsfxsize=12cm
\epsffile[20 0 600 450] {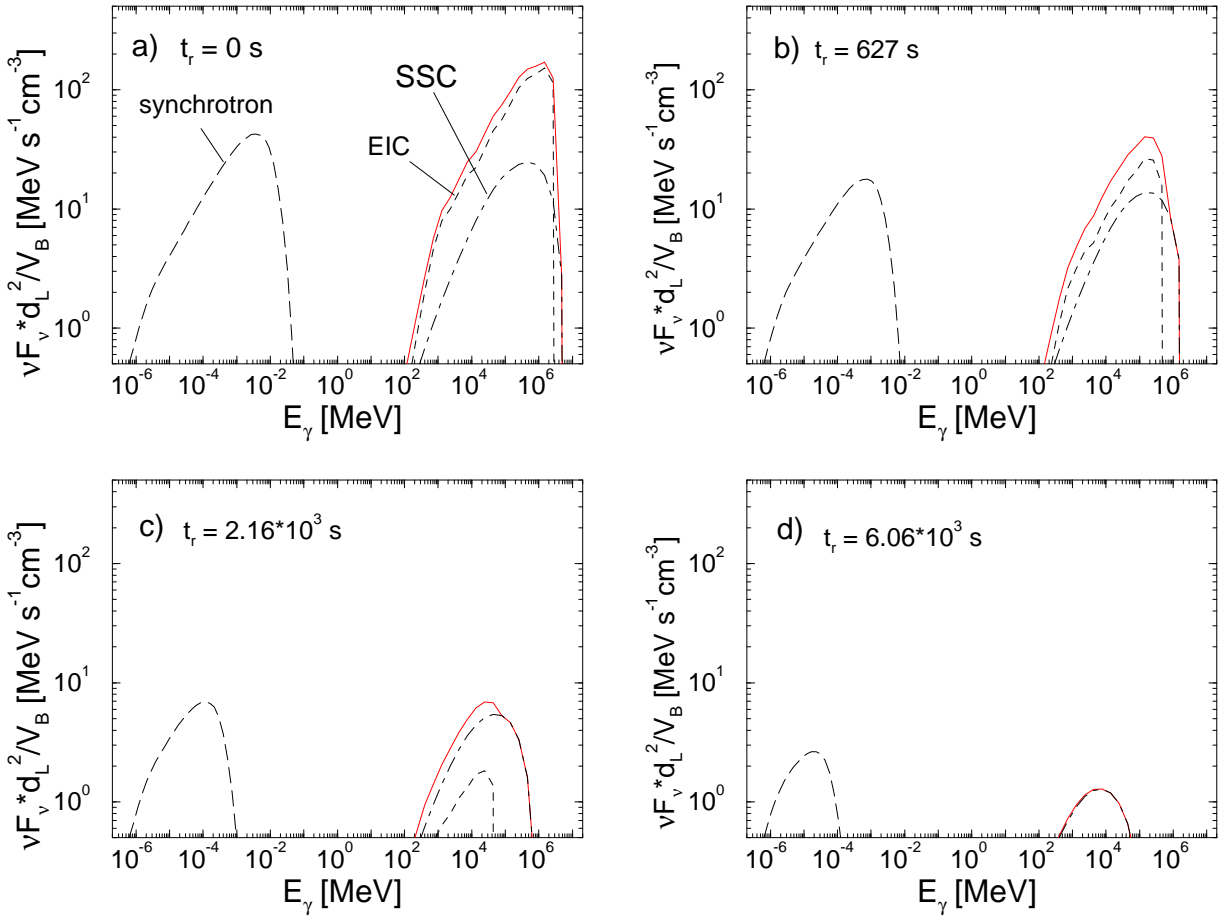} }
\caption[]{Instantaneous $\gamma$-ray spectra from the  blob of
simulation 2. Parameters: $\gamma_1 = 6 \cdot 10^3$, $\gamma_2 
= 5 \cdot 10^5$, $s = 2.0$, $R_B = 8 \cdot 10^{15}$ cm, $n = 0.2$ 
cm$^{-3}$, $B_0 = 0.07$, $B \propto z^{-0.1}$, $L_{\rm disk} = 10^{43}
\, {\rm erg \, s}^{-1}$, $M_{BH} = 10^6 \, M_{\odot}$,
 $z_i = 10^{-2}$ pc, $\Gamma = 20$, $\theta = 2^0$.
Total emission: solid; IC scattering of accretion disk 
radiation: dashed; SSC radiation: dot-dashed; synchrotron 
radiation: long dashed. (all quantities in the observer's frame) }
\end{figure*}

If a second mechanism contributes to the cooling 
of the radiating pair population, the resulting external
inverse-Compton photon spectra do not significantly
differ from the case of a pure EIC model. Equally
the assumption of a lower cutoff does not have 
an important effect on the high-energy $\gamma$-ray 
spectrum. The hard X-ray to soft $\gamma$-ray
spectrum from EIC becomes harder with increasing
lower cutoff, implying that in case of a high
lower cutoff this part of the photon spectrum
is always dominated by SSC.

Time-averaged SSC-dominated $\gamma$-ray emission 
can not account for strong spectral breaks $\Delta
\alpha \ugr 0.5$ in the MeV range as observed in
several FSRQs, even if a high lower cut-off in the
particle in the initial particle distributions is
assumed. This fact is well illustrated by Figs. 8
and 10. In contrast, such a cut-off can produce
sharp spectral breaks ($\Delta\alpha > 0.5$) in the
time-averaged EIC spectrum.

\begin{figure}
\rotate[r] {
\epsfxsize=5.8cm
\epsffile[100 20 600 50] {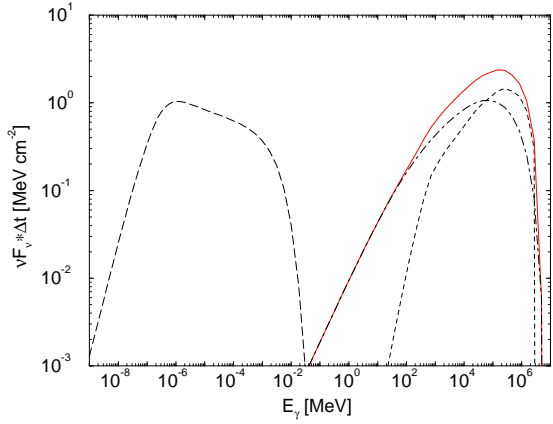} }
\caption[]{Time-integrated $\gamma$-ray spectrum from the blob
illustrated in Fig. 9 (total $\gamma$-ray spectrum: solid; external IC: 
dashed; SSC: dot-dashed; synchrotron: long dashed) }
\end{figure}

We find that even in the case of an accretion disk of
very low luminosity and injection of the pairs occurring 
relatively high above the disk EIC scattering will give
a significant contribution to the photon spectrum in the
GeV -- TeV range (see Fig. 10). If our assumption of a 
two-component $\gamma$-ray spectrum with neither SSC nor 
EIC being negligible is correct, we conclude that the 
intrinsic GeV -- TeV spectrum of blazars will in general 
not be a simple power-law extrapolation of the EGRET spectrum. 

The particle density plays a central role for the
broadband emission from ultrarelativistic plasma blobs.
The $\gamma$-ray spectra of blazars can well be
reproduced by dense blobs ($n \gg 10^4 \,
{\rm cm}^{-3}$). But in this case, escape of high-energy
$\gamma$-rays without significant $\gamma$-$\gamma$
absorption implies the presence of a very low
magnetic field and a very weak synchrotron component.
This problem is very severe for high-frequency peaked
BL Lacs (as Mrk~421) where the $\gamma$-ray component
can extend up to TeV energies at which they can be 
absorbed very efficiently by the synchrotron component
if one assumes that both components are produced in
the same volume.

In the case of very high density ($n \ugr 10^7$ cm$^{-3}$)
the cut-off to lower frequencies in the synchrotron spectra
is not caused by synchrotron-self absorption but by the 
Razin-Tsytovich effect. The same effect would also suppress
synchrotron cooling below the Razin-Tsytovich frequency
(Eq. [25]) which could consequently lead to a storage of
the kinetic energy in pairs as long as the magnetic field
does not change dramatically and the jet remains well
collimated. As mentioned above, this effect can only
be important if the synchrotron component (also above
the Razin-Tsytovich frequency) is very low compared
to the $\gamma$-ray component during a $\gamma$-ray
outburst.

Then, a time lag between a $\gamma$-ray flare and the associated 
radio outburst can give information about the geometry of the jet 
structure. We suggest that this fact could be a major reason for
the difference between FR~I and FR~II radio galaxies: If the
jet is well-collimated over kpc scales, the Razin-Tsytovich 
effect prevents efficient cooling of the relativistic pairs 
to Lorentz factors lower than the Razin Lorentz factor, and
the kinetic energy of the jet material can only be released
in the radio lobes of FR~II galaxies. If collimation is
inefficient and the jet widens up rapidly after injection, 
its kinetic energy can be dissipated at short distances from
the central engine, which would result in an FR~I structure.

\section{Model fits to Mrk~421}

We can now use our code and the results summarized in 
the last section to reproduce the broadband spectrum 
of the first extragalactic object significantly detected 
in TeV $\gamma$-rays, Mrk~421. Mrk~421 is a BL Lac object at 
$z = 0.031$ which exhibited a prominent TeV flare in May 1994 
(Kerrick et al., 1995). In this flaring phase, a flux of 
$(2.1 \pm 0.3) \cdot 10^{-10}$ photons cm$^{-2}$ s$^{-1}$ above 
250~GeV has been found. We assume $H_0 = 75 \, {\rm km \, 
(s \cdot Mpc)}^{-1}$, account for absorption by the 
intergalactic infrared background radiation (IIBR) 
using the model fits by (Stecker \& de Jager 1996) 
which do not differ very much from each other for photons of 
energies $\ukl 1$ TeV and neglect absorption by accretion disk 
radiation scattered back by surrounding clouds (B\"ottcher \& Dermer 
1995).

Our first example, shown in Figs. 7 and 8, demonstrates
that jet parameters are possible where the $\gamma$-ray
emission from Mrk~421 in its quiescent state can emerge 
from a small region where light-travel time arguments 
allow for even more rapid flickering in $\gamma$-rays 
than observed recently by Gaidos et al. (1996). As 
illustrated in Fig. 11, the synchrotron emission from 
this emission region is far below the observed radio to 
optical flux from Mrk~421.

\begin{figure}
\rotate[r] {
\epsfxsize=5.8cm
\epsffile[100 20 600 50] {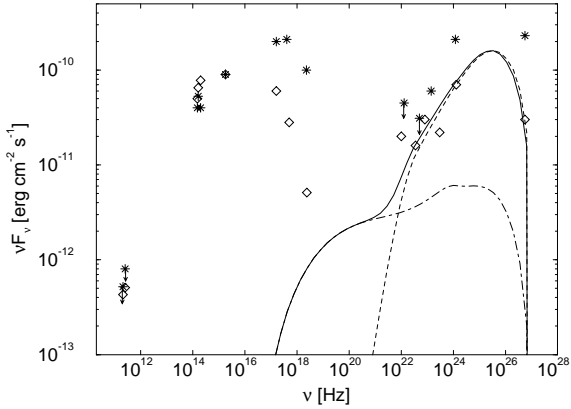} }
\caption[]{ Comparison of the time-averaged photon flux from
simulation 1 (Figs. 7, 8) to the broadband spectrum of Mrk~421 
in its quiescent state, indicated by squares (data from 
Macomb et al. [1996]) }
\end{figure}

A successful fit to the broadband spectrum of Mrk~421 in
its quiescent state can be achieved with a very low density,
as chosen for our second example, illustrated in Figs. 9 and
10. This is shown in Fig. 12.

\begin{figure}
\rotate[r] {
\epsfxsize=5.8cm
\epsffile[100 20 600 50] {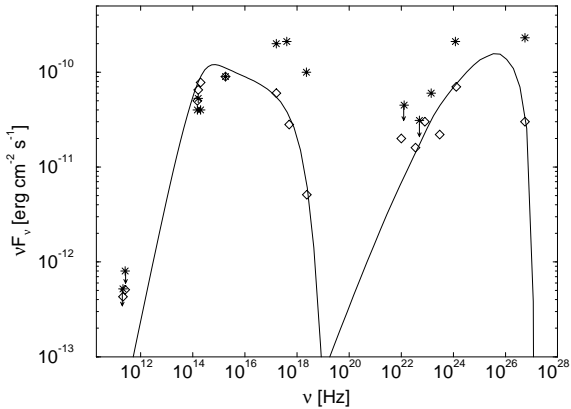} }
\caption[]{ Comparison of the time-averaged photon flux from
simulation 2 (Figs. 9, 10) to the broadband spectrum of Mrk~421 in its
quiescent state }
\end{figure}

The flaring state of Mrk~421 can result from an
increase of the maximum Lorentz factors of the
pairs which could be related to some increase
of energy input in hydromagnetic turbulences
accelerating the primary particles which can
plausibly also imply an increase in the magnetic
field. Increasing the cut-off energies to $\gamma_1
= 8 \cdot 10^3$, $\gamma_2 = 7 \cdot 10^5$ and the magnetic
field to $B = 0.08$ G while the other parameters
remain the same is in our model calculation for the
low state (see caption of Fig. 9) yields an acceptable 
fit to the flaring state of Mrk~421, as shown in 
Fig. 13. Here, we integrated over an observing time 
of $\sim 2 \cdot 10^3$ s, after which the the flux 
from this blob is far below the quiescent one.

\begin{figure}
\rotate[r] {
\epsfxsize=5.8cm
\epsffile[100 20 600 50] {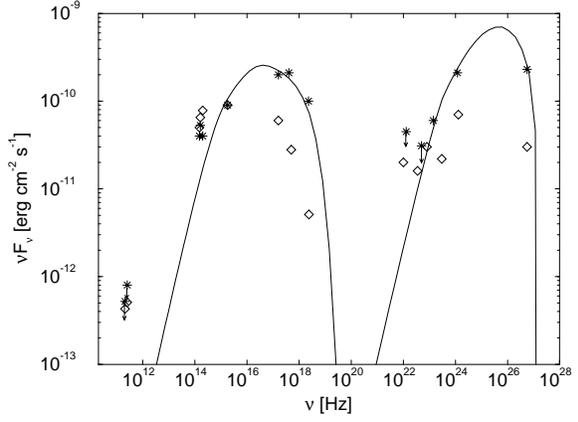} }
\caption[]{ Model fit to the broadband spectrum of Mrk~421 in its
high state (stars); $\gamma_1 = 8 \cdot 10^3$, $\gamma_2 = 5 \cdot
10^5$, $B_0 = 0.08$ G; other parameters as given in the caption
of Fig. 9 }
\end{figure}

Petry et al. (1996) found a spectral index of the differential 
spectrum of Mrk~421 above 1 TeV of $3.6 \pm 1.0$ which translates
to a spectral index of the integral spectrum of $2.6 \pm 1.0$.
In Figure 14 we demonstrate that our model fit to the low state
of Mrk~421 is in agreement with this measurement. Above 1 TeV, 
absorption by the IIBR becomes important. In order to account
for this, we used the model fits by Stecker \& de Jager (1996).
The figure demonstrates that both fits result in very similar
absorbed spectra. The HEGRA flux value is obtained in averaging 
over long observing times compared to the variability of Mrk~421. 
It lies between the Whipple data points for quiescent and flaring 
state (included in Figs. 11 -- 13) and is well in agreement 
with a duty cycle of 5~\% between high and low TeV state 
(Petry, priv. comm.).

\begin{figure}
\rotate[r] {
\epsfxsize=5.8cm
\epsffile[100 20 600 50] {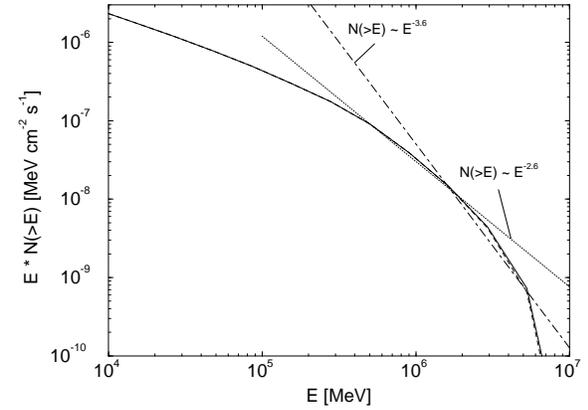} }
\caption[]{Integral photon flux resulting from simulation 2. 
The dotted and dot-dashed lines represent a power-law integral 
spectrum with spectral index 2.6 and 3.6, respectively. The
solid and dashed lines are calculated describing absorption
by the intergalactic infrared background radiation by 
Stecker \& de Jager's (1996) model fit 1 and 2, respectively }
\end{figure}

It is remarkable that the broadband fits shown here imply
an injection height of $z_i \approx 10^{-2}$ pc which is far 
outside the $\gamma$-ray photosphere for TeV photons. A high
low-energy cutoff in the electron spectrum is required in order
not to overproduce the X-ray flux. This, in turn, implies a
very low particle density in order to recover the observed
relative luminosity of the synchrotron to the SSC component.

\section{Summary and conclusions}

We gave a detailed discussion of the various radiative energy
loss mechanisms acting in relativistic pair plasmas and concentrated 
on the application to jets from active galactic nuclei. In the first 
phase after the acceleration or injection of pairs radiative cooling 
due to inverse-Compton scattering is the dominant process. Here,
Klein-Nishina effects are much more pronounced for inverse-Compton 
scattering of external radiation (from the central accretion disk) 
than for the SSC process since the synchrotron spectrum extends
over a much broader energy distribution, allowing for efficient
Thomson scattering at all particle energies.

We developed a computer code which allows to
follow the evolution of the energy distributions of
electrons/positrons self-consistently, accounting
for all Klein-Nishina effects and for the time-dependence
of the synchrotron, SSC and external radiation fields which 
contribute to the radiative cooling of the pairs and computing 
the instantaneous $\gamma$-ray production from the resulting 
pair distributions. Since the intrinsic cooling timescales
are much shorter than the time resultion of present-day
instruments, we calculated time-integrated spectra and
compared them to observations.

We found some general results from a series of different
simulations:

(a) Due to the short cooling times involved in the SSC radiation
mechanism, implying that we only see time-integrated SSC emission
leading to an overprediction of hard X-ray flux from blazars,
we favor EIC to be the dominant radiation process in $\gamma$-ray
blazars.

(b) Detailed simulations show that the EIC spectrum of a cooling
ultrarelativistic pair distribution with maximum energies
implying EIC scattering in the Klein-Nishina regime is harder 
in the case of pure Thomson scattering where we recover the 
classical results of Dermer \& Schlickeiser (1993 a).

(c) The low-energy cutoff in the synchrotron spectra from dense
pair jets is {\it not} determined by synchrotron self-absorption,
but by the Razin-Tsytovich effect. Only when the jet widens
up, radio emission at $\nu \ukl 10^{11}$ Hz can escape the jet.

\acknowledgements{We thank the anonymous referee for helpful comments.
H. Mause acknowledges financial support by the DARA (50 OR 9301 1).}


\begin{thebibliography}{}

\bibitem[14]{1990} Achatz, U., Lesch, H. \& Schlickeiser, R., 1990,
A\&A 233, 391

\bibitem[]{1993} Achatz, U. \& Schlickeiser, R., 1993, A\&A 274,
165

\bibitem[15]{1968} Alexanian, M., 1968, Phys. Rev. {\bf 165}, Nr. 1,
253

\bibitem[]{1985} Babadzhanyants, M. K. \& Belokon, E. T., 1985,
Astrofizika 23, 459

\bibitem[]{1995} Barthel, P. D., et al., 1995, ApJ 444, L21

\bibitem[7]{1967} Ba\u\i er, V. N., Fadin, V. S. \& Khoze, V. A., 1967,
Sov. Phys. JETP 24, Nr. 4, 760

\bibitem[]{1995} Becker, P. A. \& Kafatos, M., 1995, ApJ 453, 83

\bibitem[]{1995} B\"ottcher, M. \& Dermer, C. D., 1995, A\&A 302, 37

\bibitem[10]{1995} B\"ottcher, M. \& Schlickeiser, R., 1995, A\&A 302, L17

\bibitem[]{1997} B\"ottcher, M. \& Schlickeiser, R., 1997, A\&A, submitted

\bibitem[]{1992} Campeanu, A. \& Schlickeiser, R., 1992, A\&A 
263, 413

\bibitem[]{1986} Crusius, A. \& Schlickeiser, R., 1986, A\&A 
{\bf 164}, L16

\bibitem[]{1988} Crusius, A. \& Schlickeiser, R., 1988, A\&A 
195, 327

\bibitem[16]{1984} Dermer, C. D., 1984, ApJ 280, 328

\bibitem[]{1995} Dermer, C. D. \& Gehrels, N., 1995, ApJ 447, 103

\bibitem[]{1996} Dermer, C. D., Miller, J. A. \& Li, H., 1996, ApJ
456, 106

\bibitem[9]{1993} Dermer, C. D. \& Schlickeiser, R., 1993 a, ApJ
416, 458

\bibitem[]{1993} Dermer, C. D. \& Schlickeiser, R., 1993 b, Proc. of the
XXIII ICRC, Vol. I, 156

\bibitem[]{1995} Dermer, C. D., Sturner, S. J., \& Schlickeiser, R.,
1997, ApJS, in press

\bibitem[]{1993} Fichtel, C. E. et al., 1993, in: Compton
Gamma-Ray Observatory, AIP Conf. Proc. {\bf 280}, ed. M.
Friedlander, N. Gehrels \& D. J. Macomb, p. 461

\bibitem[]{1996} Gaidos, J. A., et al., 1996, Nature 383, 319

\bibitem[]{1979} Galeev, A. A., et al., 1979, ApJ 229, 318

\bibitem[]{1992} Hartman, D., et al., 1992, ApJ 385, L1

\bibitem[]{1996} Hartman, R. C., et al., 1996, ApJ 461, 698

\bibitem[4]{1975} Haug, E., 1975, Zeitschrift f. Naturforschung
30 a, 1099

\bibitem[5]{1985} Haug, E., 1985 (a), Phys. Rev. D 31, 2120

\bibitem[6]{1985} Haug, E., 1985 (b), A\&A 148, 386

\bibitem[]{1993} Henri, G., Pelletier, G. \& Roland, J., 1993, ApJ 404, 
L41

\bibitem[3]{1967} Jauch, J. M., Rohrlich, F., 1976, ``The Theory of
Photons and Electrons'', Springer, New York

\bibitem[]{1962} Kardashev, N. S., 1962, Sov. Astron. Journal 6,
317

\bibitem[]{1995} Kerrick, A. D., et al., 1995, ApJ Letters, 438, 
L59

\bibitem[]{1994} Kirk, J. G. \& Mastichiadis, A., 1992, Nature
360, 135


\bibitem[]{1987} Lee, M., A. \& Ip, W. H., 1987, JGR 92, 11041

\bibitem[]{1989} Lesch, H., Crusius, A. \& Schlickeiser, R., 1989,
A\&A 209, 427

\bibitem[]{1987} Lesch, H. \& Schlickeiser, R., 1987, A\&A 179,
93

\bibitem[]{1969} Lynden-Bell, D., 1969, Nature 223, 690

\bibitem[]{1995} Macomb, D. J. et al., 1995, ApJ 449, L99

\bibitem[]{1995} Macomb, D. J. et al., 1995, ApJ 459, L111

\bibitem[]{1992} Mannheim, K. \& Biermann, P. L., 1992, A\&A 253,
L21

\bibitem[]{1995} Mastichiadis, A. \& Kirk, J. G., 1995, A\&A 295,
613

\bibitem[]{1969} McKenzie, J. K. \& Westphal, K. O., 1969, Planet Space
Sci. 17, 1029

\bibitem[]{1996} Petry, D., Bradbury, S. M., Konopelko, A. et al., 1996,
A\&A 311, L13

\bibitem[]{1992} Pohl, M., Reich, W. \& Schlickeiser, R., 1992, A\&A
262, 441

\bibitem[]{1995} Pohl, M., Reich, W., Krichbaum, T. P., et al., 1995,
A\&A 303, 383

\bibitem[]{1992} Punch, M., et al., 1992, Nature 358, 477

\bibitem[]{1995} Quinn, J., 1995, IAU Circ. No. 6169

\bibitem[]{1984} Rees, M., J., 1984, ARA\&A 22, 471

\bibitem[8]{1979} Rybicki, G. B., Lightman, A. P., 1979, ``Radiative
processes in astrophysics'', John Wiley \& Sons, New York

\bibitem[]{1964} Salpeter, E., 1964, ApJ 140, 796

\bibitem[]{1984} Schlickeiser, R., 1984, A\&A 136, 227

\bibitem[]{1994} Schlickeiser, R., 1994, ApJ Suppl. 90, 929

\bibitem[]{1992} Schlickeiser, R. \& Achatz, U., 1992, in: Extragalactic
Radio Sources --- From Beams to Jets, eds. J. Roland, H. Sol, G. 
Pelletier, Cambridge University Press, Cambridge, p. 214

\bibitem[]{1992} Schlickeiser, R. \& Achatz, U., 1993: J. Plasma Phys.
49, 63

\bibitem[]{1989} Schlickeiser, R., Crusius, A., 1989, IEE transactions
on plasma science, Vol. 17, No. 2, 245

\bibitem[18]{1973} Shakura, N. I., Sunyaev, R. A., 1973, A\&A
24, 337

\bibitem[]{1987} Sikora, M., Kirk, J. T., Begelman, M. C. et al.,
1987, ApJ 320, L81

\bibitem[]{1996} Sikora, M., Madejski, G., Moderski, R. et al., 1996,
ApJ, submitted

\bibitem[17]{1995} Skibo, J. G., Dermer, C. D., Ramaty, R. \&
McKinley, J. M., 1995, ApJ 446, 86

\bibitem[]{1996} Stecker, F. W., de Jager, O. C., 1996, ApJ, in press

\bibitem[11]{1982} Svensson, R., 1982, ApJ {\bf 258}, 321

\end{thebibliography}
\end{document}